\RequirePackage{fix-cm}

\documentclass[natbib,smallextended]{svjour3}       

\smartqed  
\usepackage{graphicx}
 \journalname{my journal}
\begin{document}

\title{On the radio spectra of supernova remnants
}


\author{Dejan Uro{\v s}evi{\' c}
}


\institute{D. Uro{\v s}evi{\' c} \at
              Department of Astronomy, Faculty of Mathematics, University of Belgrade, Studentski trg 16, 11000 Belgrade, Serbia \\
              Tel.: +381-11-2027827\\
              Fax: +381-11-2630151\\
              \email{dejanu@math.rs}           
           }

\date{Received: date / Accepted: date}

\maketitle

\begin{abstract}

The theoretical fundamentals of formation of the supernova remnant (SNR) continuum radio spectra are presented in this review. Mainly based on the Fermi 1 theory - also known as diffuse shock acceleration (DSA) - the different shapes (linear or curved in log-log scale) of SNR radio spectra are predicted for both young and evolved SNRs. On the other hand, some particular forms of spectra of older SNRs can be predicted by including the additional processes such as Fermi 2 acceleration mechanism or thermal bremsstrahlung radiation. Also, all of these theoretically predicted forms of radio spectra are compared with real spectra obtained from observations. Finally this review can represent some kind of "atlas" with initial patterns for the different kinds of SNR radio spectra -- it should be helpful for radio astronomers in their interpretation of the observed radio spectra.

\end{abstract}

\section{Introduction}
\label{intro}

The synchrotron (magneto-bremsstrahlung) mechanism represents the main production mechanism of radio waves in astrophysical sources. It is radiation produced by ultra-relativistic electrons gyrating in a magnetic field. The two fundamental ingredients of the synchrotron process - magnetic field and
high-energy electrons - should provide a reservoir of energy which can be transformed to synchrotron radiation. The synchrotron electrons have helicoidal paths around the field lines and their radiation is highly beamed in the direction of velocity vectors. The magnetic field is always embedded in the interstellar medium (ISM), but electrons should be previously accelerated to the ultra-relativistic velocities to be capable of radiating by the synchrotron process. Due to this, we have to know very thoroughly on which way electrons can be accelerated to ultra-relativistic energies. The most efficient process of particle acceleration is the so-called diffuse shock acceleration (DSA)(e.g. Bell 1978a, Blandford and Ostriker 1978). Also known as the Fermi 1 acceleration mechanism, electrons gain energy as the result of multiple crossing of a charged particle through a shock wave. If an astrophysical object contains some form of shock, we can expect acceleration of charged particles from the medium around the shock. Although heavier particles (such as protons and ions) can be accelerated very efficiently to $\sim10^{15}$ eV, electrons can also be accelerated to the ultra-relativistic energies ($\sim10^{12}$ eV) by DSA mechanism -- this lower maximal energy of electrons originates from the very rapidly energy loses induced by the synchrotron radiation. If a shock wave propagates through ISM, the electrons from ISM should be accelerated and be captured by the ISM magnetic field, thus generating synchrotron radiation. The supernova remnants (SNRs) (which can be seen in our Galaxy and in nearby galaxies),  active galactic nuclei (AGNs) with associated jets and lobes, pulsars and pulsar wind nebulae are strong synchrotron emitters. All of these objects have their own characteristic forms of synchrotron radio spectra. The shapes of these spectra essentially depend on particle acceleration processes. In this review, the different forms of SNR spectra (linear or curved in log-log scale, hereafter linear or curved, respectively) for young and evolved SNRs are predicted and compared with spectra obtained from radio observations. In Section 2 the DSA mechanism and a way of formation of synchrotron radio spectra (as a direct consequence of the acceleration mechanisms) are briefly presented. The theoretically predicted radio spectra of young and evolved SNRs are reviewed in Sections 3 and 4, respectively. In Section 5, the previously theoretically predicted shapes of radio spectra are compared with the observationally obtained radio spectra. The summary of this review is given in Section 6.

\section{DSA mechanism and synchrotron radiation}

\label{sec:1}

\subsection{The brief review of the DSA fundamentals}

Fermi proposed in his crucial paper (Fermi 1949) the way of acceleration of charged particles to the energy of cosmic rays (CRs) and derived that the energy gain in one collision between a particle and the magnetic perturbation to be $\propto (u/v)^2$, where $u$ is the velocity of magnetic perturbation, and $v$ is the velocity of high-energy particle. In the interstellar case, the mechanism proposed by Fermi is not very
efficient but the fact that particles can gain energy in collisions with the irregularities
of the magnetic field is at the basis of all acceleration theories.
Later, when the Fermi model was established as the ideal starting point, this original mechanism of particle acceleration was called the second order Fermi acceleration mechanism (also known as {\it Fermi 2 acceleration}). On the other hand, DSA is the first order Fermi acceleration mechanism (also known as {\it Fermi 1 acceleration}) where during the collision between a charged particle and the magnetic perturbation (in the case of DSA the magnetic perturbation is represented by a magnetized collisionless shock discontinuity)\footnote{The magnetic perturbations (from which a high-energy charge particle can be reflected), are connected with turbulent fluid motion in the downstream region. Fast particles are prevented from streaming away upstream of a shock front by the scattering of Alfv{\'e}n waves which they themselves generate and essentially represent magnetic perturbations in the upstream region.}, the gain of particle energy is $\propto u/v$ (e.g. Bell 1978a).
In the original version of the Fermi acceleration (Fermi 1949) + and $-$ signs in front of linear part $(u/v)$ correspond, respectively, to the approaching
or receding motion of the magnetic mirror, and hence to a gain or a loss
of energy. Due to this, the linear dependence disappears from corresponding equations. Since the direct collisions (approaching) are statistically more numerous
than the inverse ones (receding), due to the relative velocities, there is a net energy
gain for the particle, whose energy increases continuously due to the accumulated
collisions, and this gain is represented by survived second order dependence $(u/v)^2$.
On the other hand, the net energy gain in an acceleration process would be substantially higher if there were only direct (head-on) collisions. It provides that
increase is $\Delta E/E \propto u/v$, that is, first order in $u/v$ and, appropriately, this is referred to
as first-order Fermi acceleration (e.g. Longair 2000, Lequeux 2005). Due to this dependence, when only one collision is considered, DSA is more efficient process than Fermi 2 mechanisms.
Further, DSA theory is based on multiple transition of one charged particle through the shock discontinuity from upstream to downstream region and vice-versa. In every passage (head-on) across the shock, independent from which side of shock the passage occurs, the test particle gains energy. This process provides high efficiency of particle acceleration\footnote{The relation between the characteristic acceleration times to energy $E$ for DSA and Fermi 2 mechanisms is given by $t_{\rm acc}({\rm DSA})\propto M_{\rm A2}^{-2}t_{\rm acc}({\rm Fermi\, 2})$ (Reynolds 2008), where $M_{\rm A2}=u_2/v_{\rm A}$ is the Alfv{\'e}n Mach number of the downstream flow; $u_2$ is the downstream fluid velocity and $v_{\rm A}$ is the Alfv{\'e}n speed, defined by $v_{\rm A}=B/(\mu_0 \varrho)^{1/2}$, where $B$ is the magnetic field strength, $\mu_0$ is the magnetic permeability of the vacuum and $\varrho$ is the mass density.}. The basic assumption in this theory (the so-called test particle theory) is that the particle before the actual start of DSA mechanism has very high velocity $v\gg u$. Of course, DSA model is generalized to include lower starting velocities of test particles (e.g. Bell 1978b). The very important result of DSA theory is the prediction of the distribution function for high-energy particles (or energy spectrum). This theory predicted a power-law energy spectrum of accelerated particles in the form $N(E)\propto E^{-\gamma}$, where $N(E)$ is the number of particles per unit volume at energy $E$ and $\gamma$ is the energy index. Furthermore this theory predicted a value of $\gamma=2$ (Bell 1978a, Blandford and Ostriker 1978) for acceleration of particles at the strong shock waves with the compression ratio $r=u_1/u_2=4$, where $u_1$ and $u_2$ are the upstream and downstream fluid velocities with reference to shock front, respectively. Here, only the fundamentals of linear DSA are presented. In the next sections, the upgraded versions of DSA theory will be presented, where necessary, for the explanation of different forms of radio spectra.

\subsection{The brief review of the Synchrotron radiation fundamentals}

Electrons are charged particles which can be accelerated to ultra-relativistic energies by DSA mechanism. Also, their energy spectrum at high energies is a power-law with the same value of the energy index $\gamma=2$, as it is derived for protons and heavier ions (Bell 1978b). Due to this, DSA mechanism can provide an energy contingent which can be devoted, together with energy contained in the magnetic field, to the synchrotron emission. The synchrotron emission of one electron is continuum emission with its maximum near the so-called critical frequency (Ginzburg and Syrovatskii 1965). The frequency of maximal emission for an electron with energy $E$ in the magnetic field of strength $B$, with a component $B_\perp$ orthogonal to the velocity vector, is defined by the expression: $\nu/{\rm MHz}=16.0(E/{\rm GeV})^2(B_\perp/\mu{\rm G})$ (Cummings 1973). By using this formula we can calculate the kinetic energies of electrons which radiate by the synchrotron mechanism, and their maxima of emitting energy correspond to the limiting frequencies of the radio domain (10MHz - 100GHz). The corresponding energies of electrons are 800MeV (at 10 MHz) and 80GeV (at 100 GHz), for $B_\perp=1\mu$G - the characteristic value of ISM magnetic field. These energies are obviously ultra-relativistic. For example, if we consider the SNR emission, the electrons which mainly produce radiation by the synchrotron mechanism at 1 GHz have energy $\sim800$MeV\footnote{The value for the magnetic field used here is 100$\mu$G. It is standard value for the compressed and amplified ISM magnetic field by the strong shock wave of an SNR (Arbutina et al. 2012, 2013).}. It is clear that the synchrotron radiation at one frequency arises mainly from electrons which have approximately same kinetic energy. On the other hand, the energy spectrum of relativistic electrons which is in the form of power-law has to be transformed in the power-law radio spectrum. The energy index from the energy spectrum can be transformed in the so-called spectral index of the radio spectrum ($\alpha$), by the simple linear equation $\gamma=2\alpha+1$. Finally, a radio spectrum may be characterized by the power-law form $S_\nu\propto\nu^{-\alpha}$, where $S_\nu$ is the radio flux density at frequency $\nu$. Obviously, the value for spectral index derived directly from DSA theory is $\alpha=0.5$. As measured by observations, the large majority of radio spectra of SNRs (approximately 20\% of all observed Galactic SNRs) have $\alpha$=0.5!\footnote{SNR spectral indices are mainly between 0.2 and 0.8.} (e.g. see available catalogues of Galactic SNR by Guseinov et al. (2003, 2004a,b) and Green (2009)). It is excellent confirmation that predictions of the DSA theory are correct. The theoretical fundamentals of formation of the curved SNR radio spectra and linear spectra for which $\alpha\neq0.5$ will be presented in the next sections.

\section{Radio spectra of young SNRs}

\subsection{The power-law (linear in log-log scale) radio spectra of young SNRs}

In this review, the young SNRs are objects where the particle acceleration is very efficient. It mainly corresponds to free-expansion and especially early Sedov phases of SNR evolution (Woltjer 1972, Berezhko and V{\"o}lk 2004, Pavlovi{\' c} et al. 2013). Observations show that young SNRs have steep radio spectral indices, around 0.7 (see Green 2009). It is significantly steeper than the expected $\alpha=0.5$, derived from test particle DSA. The theory proposed by Bell's (1978a,b) predicted steep spectral indices for older SNRs, when the shock wave loses significant part of its energy and particle acceleration is less efficient. Contrary to this prediction we obtain from observations the steep indices for young SNRs, for which the particle acceleration should be highly efficient because of the very fast shock waves (between a few thousand and a few tens of thousand km/s), and for which harder spectra should be expected (shallower spectral indices around 0.5). The explanation of this controversy was given in paper of Bell et al. (2011). In the original Bell's theory from 1978, it was assumed that the magnetic field direction is parallel to the shock normal. This assumption provides efficient diffusion from upstream to downstream region and efficient DSA. Due to this, the resulting spectra of high-energy particles should be harder. Bell et al. (2011) showed that young SNRs with the quasi-perpendicular orientation of the magnetic field should have steeper spectral indices. The magnetic field might be mainly quasi-perpendicular in the event of strong magnetic field amplification\footnote{The magnetic field amplification is driven by the non-linear effects, when pressure of CRs, in addition to the gas pressure, represents a significant part of the total pressure (Bell 2004). A detailed description of the non-linear DSA will be presented later in this review.} in the precursor driven by the cosmic ray (CR) streaming (Bell 2004, 2005). Field amplification is induced by $\mathbf{j}_{\rm CR}\times\mathbf{B}$, where $\mathbf{j}_{\rm CR}$ is electric current carried by CR streaming and $\mathbf{B}$ is the local (small scale) magnetic field. The background plasma is in motion perpendicular to CR current and therefore predominantly perpendicular to the
shock normal since $\mathbf{j}_{\rm CR}$ is directed away from the shock. Moving of the background plasma produces current in opposite direction to CR current. This is necessary process to provide electrical quasi-neutrality in the system. Finally, the upstream fluid motion stretches the magnetic field orthogonal to the shock normal, producing a quasi-perpendicular orientation of the magnetic field. Briefly, the field amplification leads predominantly to a quasi-perpendicular magnetic field geometry, and this geometry causes the steepening of the spectral index (Fig. 1). This is not contradicted by polarimetric observations
which suggest a predominantly radial large scale magnetic fields which are located mainly in downstream regions of SNR shocks.

\begin{figure}[h]
  \includegraphics{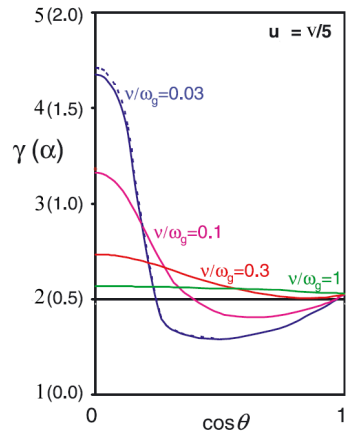}
  \centering
\caption{Spectral indices at shock velocity $v/5$ and four CR scattering rates ($\nu/\omega_{\rm g}$ = 0.03, 0.1, 0.3 and 1.0) - Figure 1, panel (a), in Bell et al. (2011). $\theta$ is the shock obliquity. At a perpendicular shock $\cos\theta = 0$. At a parallel shock $\cos\theta = 1$. The dotted line next to the curve for  $\nu/\omega_{\rm g}$= 0.03 is a calculation with finer spatial resolution. For more details see Bell et al. (2011).}
\end{figure}

Bell et al. (2011) gave for the first time an explanation based on the DSA theory for the observed steep spectral indices in young SNRs. This model is comprehensive and free of incorrect assumptions. The steeper spectral indices are not obtained only for the low CR scattering rates, as clearly shown in Fig. 1. Alternatively, the geometry might be preferentially quasi-perpendicular if an SNR expands into an ISM or circumstellar medium (CSM) in which a pre-existing  magnetic field is nearly parallel to the shock front.

\medskip

Jiang et al. (2013) have recently presented a new explanation for the steeper spectra of young SNRs. They emphasized that Alfv{\'e}nic drift effect can make softer spectra (steeper spectral indices). This effect represents a moving of guiding centers of the gyrating charged particles induced by Alfven waves. The Alfv{\'e}nic drift effect depends on the magnetic field because the Alfv{\'e}n Mach number is inversely proportional to the magnetic field $M_{\rm A}=u_1/v_{\rm A}\propto 1/B$. In the amplified magnetic fields by the non-linear modified shocks, the Alfv{\'e}n speed $v_{\rm A}\propto B\varrho^{-1/2}$ has to be higher and thus the Alfv{\'e}nic drift effect is stronger. Therefore, the Alfv{\'e}n Mach number has to be lower. This leads to the reduced shock compression ratio, and therefore the power-law spectrum of young SNRs should be steeper.

The explanation for spectral steepening of young SNRs given by Jiang et al. (2013) is simple and therefore very attractive. On the other hand, this model assumes the test particle case in the non-linear DSA approach. These two concepts are essentially in contradiction. For this reason, the explanation proposed by Jiang et al. (2013) should be considered with caution.

\subsection{The curved radio spectra of young SNRs}

The curved, concave-up radio spectra of young SNRs should be the result of non-linear DSA effects (e.g. Reynolds and Ellison 1992, Allen et al. 2008). The pressure of CR particles produced at the shock wave is included in fluid dynamics equations (see Drury 1983). The shock structure must be self-consistently calculated by including the CR pressure and cannot just be assumed. The modification of shock structure implies changes in the spectrum of accelerated electrons so that this is a highly non-linear problem. Basically, the upstream plasma is decelerated by the CR pressure and heated by plasma at the CR precursor region. Due to this, the effective Mach number becomes smaller at the jump of the subshock\footnote{The idealized vertical profile of the shock discontinuity is degenerated into the weak subshock discontinuity plus the wide transition region, which together represent a modified shock.}, and the compression ratio at the subshock becomes less than four ($r<4$). The total compression ratio of the entire modified shock (from far upstream to the downstream) becomes higher than four ($r>4$). This higher compression ratio appears due to the existence of high-energy (CR) particles in the transition region, with adiabatic index of $\gamma_{\rm ad}=4/3$, $(r=(\gamma_{\rm ad}+1)/(\gamma_{\rm ad}-1))$. Also, the higher compression ratio originates from loss of the shock energy by CR escaping from the transition region. It is similar situation as for the radiative shocks where photons carry off energy from the system. Due to this, very high compression ratios are possible, as is the case with radiative shocks (Eichler 1984). The change of the compression ratio through a modified shock should induce a change in the energy spectrum of accelerated electrons from that predicted by standard DSA. The energy spectrum of low energy electrons becomes softer (the radio spectrum is steeper) than that of linear DSA, because low energy electrons, with smaller Larmor radii, can penetrate only through the subshock ($r<4$). The spectrum of high energy electrons becomes harder (the radio spectrum is shallower) than that of linear DSA, because high energy electrons, with bigger Larmor radius, can pass through the entire modified shock (subshock plus CR precursor, $r>4$). Due to significant CR production, we can expect the concave-up radio spectra observed in the cases of young SNRs (Drury and V{\" o}lk 1981, Longair 2000).

This theory which represents the basis for obtaining the concave-up radio spectra of young SNRs is well established, given in logically complete form, at the same time very comprehensive and quite useful. The modified shocks are closer to the particularly strong shocks which can be detected in the ISM. The modified shocks have finite thickness, contrary to idealized shock discontinuities which are infinitely thin. Until the introduction of the modified shocks, this idealization of an infinitely thin discontinuity has been used in all previous theoretical approaches given in this review.

\section{Radio spectra of evolved SNRs}

\subsection{The linear radio spectra of evolved SNRs}

In the cases of evolved SNRs, the particle acceleration mechanisms are not efficient. Moreover, their influences on the SNR dynamics gradually decline and become insignificant relatively fast. We can generally say they are in the later phases of the Sedov phase, and in the radiative phases of SNR evolution (Woltjer 1972, McKee and Ostrkier 1977, Berezhko and V{\"o}lk 2004). The linear DSA predicts spectral indices around 0.5 and they should be steeper with SNR evolution (Bell 1978a,b). The greatest number of evolved SNRs have spectral indices in the interval $0.5\leq\alpha\leq0.6$. It is in agreement with the DSA prediction: the particle acceleration efficiency should decline for an older SNR where its associated shock is weaker, due to this the corresponding compression ratio is lower, and finally, its radio spectrum has to be steeper. On the other hand, a number of evolved shell-like SNRs have spectral indices $\alpha<0.5$.
These harder spectra can not be explained only by the DSA theory (neither linear nor non-linear). The particle acceleration mechanisms responsible for production of this flatter form of radio spectra should be the second-order Fermi acceleration mechanism coupled with DSA. Electrons are also accelerated by the second-order acceleration process, in the turbulent downstream plasma. Schlickeiser and F{\" u}rst (1989) proposed an explanation of flat SNR radio spectra by including simultaneously the energy gain from shock crossing and second-order Fermi acceleration in a low $\beta$-plasma\footnote{Plasma $\beta=p/p_{\rm mag}$, where $p$ is the gas pressure, and $p_{\rm mag}$ is the magnetic pressure.}. Ostrowski (1999) presented a modified approach to the acceleration process at a shock wave, in the presence of the second-order Fermi acceleration in the turbulent medium near the shock. An alternative explanation for the observed flat synchrotron spectra of SNRs in molecular clouds was discussed. It was shown that shocks may produce very flat cosmic ray particle spectra
in the presence of momentum diffusion. Due to this, the medium Alfv{\'e}n Mach number shocks for such SNRs can naturally lead to the flatter observed spectral indices.

The coupling of DSA and Fermi 2 mechanisms should represent the next step in the development of particle acceleration theory for all types of SNRs and other high energy astrophysical objects. It will be a demanding task in the sense of theoretical analysis. For providing comprehensive results of numerical simulations, significant computational power should be used.

\medskip

At the first glance, the flatter spectra should be expected for very old SNRs in the radiative phases of evolution. The fully radiative (isothermal) shocks feature a very high compression ratio $r=M_T^2$, where $M_T$ is the isothermal Mach number. Due to this, the spectral indices should be significantly flatter because $\alpha=3/(2(r-1))$. Here, we should emphasize that for the flat spectral indices of SNRs $(0.2<\alpha<0.5)$, the corresponding fully radiative shocks should be very weak, i.e. the values of isothermal Mach numbers should be between 2 and 3 (Fig. 2). Due to this, the detection of a fully radiative SNR with a flat spectral index is highly unrealistic.

\begin{figure}[h]
  \centering
  \includegraphics{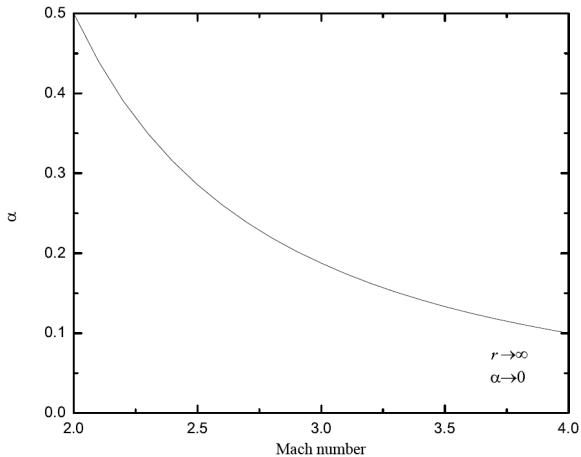}
\caption{Change of the radio spectral index as function of isothermal Mach number for the totally radiative shock of an SNR (Oni{\' c} 2013b).}
\end{figure}

For a detailed review on SNRs with flat radio spectra, see Oni{\' c} (2013b) and references therein.

\subsection{The curved radio spectra of evolved SNRs}

 A few evolved SNRs embedded in a dense ISM environment produce the curved radio spectra as a result of a significant component of the thermal bremsstrahlung radiation in their total radio emission (see e.g. Uro{\v s}evi{\' c} and Pannuti 2005, Uro{\v s}evi{\' c} et al. 2007, Oni{\' c} et al. 2012). Especially at the higher radio frequencies, the thermal bremsstrahlung component becomes significant ($S_\nu\propto\nu^{-0.1})$, and when combined with the nominal synchrotron emission detected from SNRs,
the spectrum appears flatter than expected when extrapolating from lower radio frequencies where synchrotron component is dominant ($S_\nu\propto\nu^{-\alpha})$. The thermal bremsstrahlung component manifests itself as a bend in a spectrum in the concave-up form. The necessary conditions for significant thermal emission of an object are that the radiative plasma should be at high densities and low temperatures: 1-10 ${\rm cm^{-3}}$ and 1000 - 10000 K, respectively (e.g. Uro{\v s}evi{\' c} and Pannuti 2005, Uro{\v s}evi{\' c} et al. 2007). It corresponds to evolved SNRs, especially the class of so-called mixed-morphology or also often called thermal composite \footnote{The mixed-morphology or thermal composite SNRs appear as shell like in radio, but with bright interiors due to thermal X-rays (e.g. Rho and Petre 1998, Jones et al. 1998). } SNRs, embedded in molecular cloud environment (Oni{\' c} et al. 2012). On the other hand, the high frequency thermal emission naturally leads to low-frequency thermal absorption. Due to this we should expect that this concave-up form of spectrum turnovers into the concave-down form at the low-frequency end of the radio domain (for details see Oni{\' c} et al. (2012), Oni{\' c} (2013a,b)).

This "thermally active" model is simple and intuitively logical. On the other hand, the most important drawback in this approach is related to the cooling rate of hot gas in an SNR interior. Due to this, the thermal bremsstrahlung emission is not expected from the hot interiors. It is expected from the same volume from which the synchrotron emission is generated, or from CR and radiative precursors. However this model works very well for SNRs embedded in dense ($n>1\, {\rm cm^{-3}}$ ionized, atomic and primary molecular) environment (e.g. Galactic SNR 3C396, see Oni{\' c} et al. (2012)).

\medskip

The radio spectra of evolved SNRs could be curved in the opposite direction, i.e. they appear in concave-down form. This kind of spectrum can be explained using DSA theory with the effect of synchrotron losses within the finite emission region. If the thin region near the shock discontinuity is not resolved by the telescope beam, the observed emission contains some emission from electrons which have been diffused away from the place of effective acceleration and lose a significant amount of energy via the synchrotron emission. As these losses are more severe for higher energy electrons, we expect this to steepen the observed synchrotron spectrum. For details see Heavens and Meisenheimer (1987), Longair (2000, and references therein). This concave-down form of radio spectra should correspond to very old SNRs, for which electrons have had enough time\footnote{The lifetime of an SNR, as defined by McKee and Ostriker (1977), can be up to one million years -- this timescale is comparable to the typical timescale of synchrotron loss for electrons which radiate at the highest radio frequencies.}to lose a significant amount of energy at the highest radio frequencies, and primarily to distant (mainly extragalactic) SNRs for which the limitation in telescope resolution leads to the capturing of radio emission from the sample of "exhausted" electrons.

\section{Examples of different forms of SNR radio spectra obtained from observations}

The radio spectra of SNRs obtained from observations appear predominantly in linear form but with obvious scattering in the flux densities (e.g. Gao et al. (2011), Sun et al. (2011)). Furthermore, they are generally poorly-studied with respect to number of observational data points at the frequencies available for radio observations. Also, a significant dispersion of values of the flux densities at the same frequency can be found. As a primary problem, the determination of accurate values of the integrated flux densities is a very demanding task with non-unique exiting values. Additionally, the radio spectral index variations, both with frequency and location within several Galactic SNRs were detected and described in the last three decades (for details see Oni{\' c} (2013a,b), and references therein). Due to these difficulties, at least for the majority of observed SNRs, the present knowledge of radio spectral indices as well as of the overall shapes of the SNR radio continuum spectra is far from being precise. However, some trends toward the curved spectral forms can be found in relatively recent literature.

\subsection{Young SNRs}

There is a considerable amount of observational evidence for steep spectral indices of young SNRs. The brief inspection of Galactic SNR catalogues (e.g. Guseinov et al. 2003, 2004a,b, Green 2009) shows that all historical Galactic SNRs (RCW86 - probably remnant of SN185, SN1006, Tycho SNR, Kepler SNR, Cas A, G1.9+0.3)\footnote{Crab nebula and SNR 3C58 represent particular type of SNR, the so-called pulsar wind nebulae, which are not target of this review. Contrary to shell-like, composite and mixed-morphology SNRs which are target of this review, the pulsar wind nebulae typically have flat spectral indices between 0.0 and 0.3 (e.g. Reynolds et al 2012).} have steep spectral indices between 0.6 and 0.8. The modifications of DSA theory described in this review, in Subsection 3.1, can explain these apparent steeper linear spectral forms of young SNRs.

\begin{figure}
  \centering
  \includegraphics{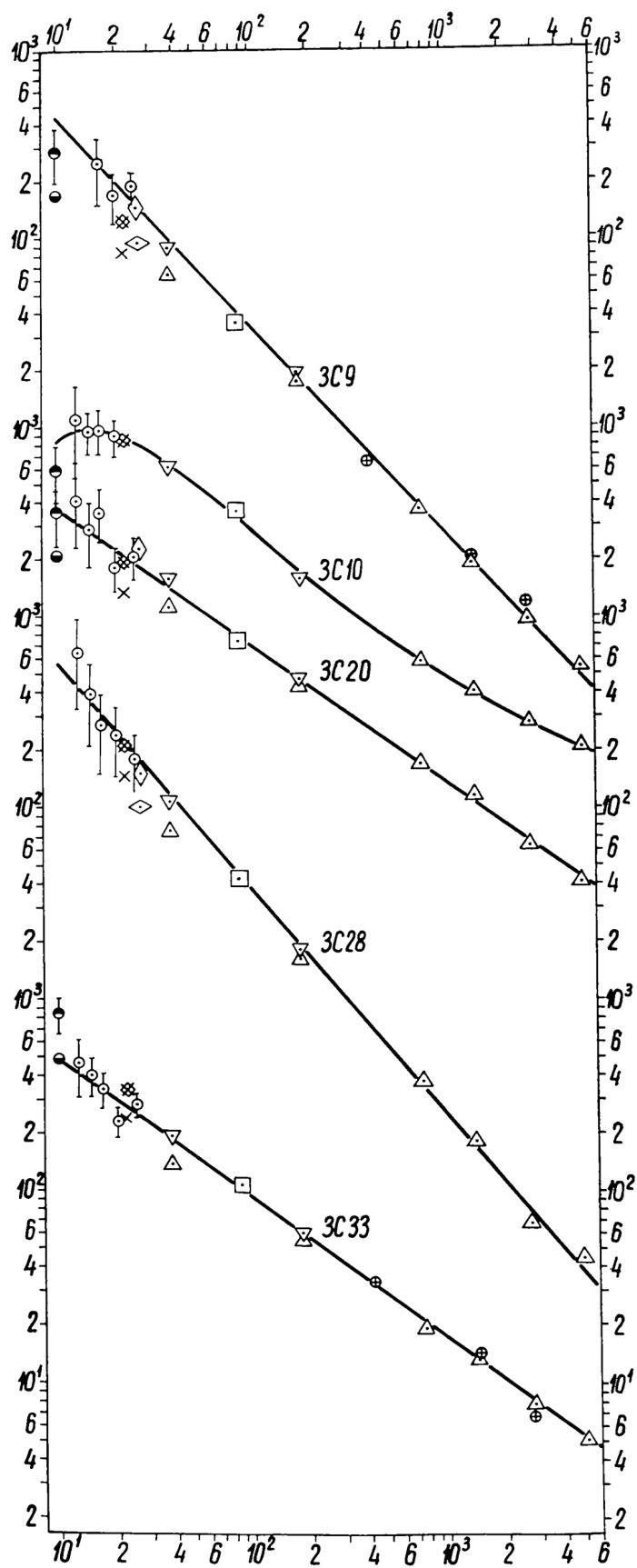}
\caption{The spectra of five discrete radio sources in 10 MHz - 5 GHz frequency range (Braude et al. 1970). The spectrum of Tycho SNR (3C10) is shown - evidently in tentative concave-up form.}
\end{figure}

Besides spectra with linear profiles, some of the historical Galactic SNRs show tentatively curved concave-up spectra (hardening to higher energy) predicted by non-linear DSA (this review, Subsection 3.2). Braude et al. (1970) detected mildly concave-up form of Tycho SNR spectrum in the range between 38 MHz and 5 GHz (Fig. 3). It was the first indication of concave-up curvature for an SNR spectrum. After that, Roger et al. (1973) once again confirmed detection of the same spectral form of Tycho SNR. The explanation of this phenomenon was not examined until Reynolds and Ellison (1992) showed that the spectral curvature observed in Tycho and Kepler SNRs should be explained by the non-linear DSA effects. They presented the first model for synchrotron spectra calculated with the self-consistent, modified shock model by the non-linear DSA and compared them with the observed radio spectra of Tycho and Kepler SNRs (Fig. 4), finding reasonable agreement. Additionally, they predicted that if inconsistencies in the total radio spectrum of SN 1006 SNR were resolved, their curved models would fit just as well there. Jones et al. (2003) gave  observational evidence that the radio spectrum of Cas A should be concave-up due to their positive identification of infrared synchrotron radiation from this SNR. Their flux density at 2.2 $\mu$m, after
correction for extinction, falls significantly above the power-law extrapolation from the radio.
If the curvature were constant across this full range of wavelengths, then the local spectral index would change from approximately 0.75 at cm wavelengths to approximately 0.5 at 2.2 $\mu$m. Similar as in previous case, Vink et al. (2006) obtained the concave-up form of spectrum for RCW86 SNR, but using X-ray data. They analyzed the X-ray synchrotron emission from this SNR and concluded that concave-up spectrum is necessary in order to properly fit radio and X-ray data together. Finally, Allen et al. (2008) showed the evidence of a concave-up synchrotron spectrum of SN 1006, demonstrating for the first time that the synchrotron spectrum of SN 1006 seems to flatten with increasing energy. The evidence that the synchrotron spectra (from radio to X-rays) of SN 1006 are curved can be interpreted as evidence of curvature in the GeV-to-TeV electron spectra. All previously described concave-up spectra obtained from observations present the evidence of curved electron spectra and suggest that cosmic rays are not ''test'' particles. The cosmic-ray pressure at the shock is large enough to modify the structure of the shock.

\begin{figure}[h]
  \centering
  \includegraphics{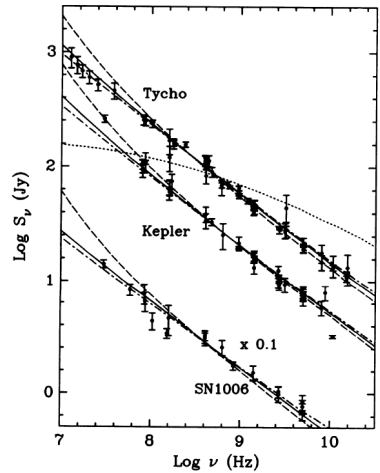}
\caption{Slightly concave-up spectra of three historical SNRs (Reynolds and Ellison 1992).}
\end{figure}

\subsection{Evolved SNRs}

The largest fraction of observed SNRs is characterized by spectral indices in the range $\alpha\in(0.5,0.55)$. These values were predicted by the test particle DSA theory. On the other hand a significant number of observed SNRs have flatter indices $\alpha<0.5$ (this review, Subsection 4.1). There are several explanations of the harder radio spectra of SNRs: significant contribution to the observed emission of the second-order Fermi mechanism, high compression ratio ($> 4$), contribution of secondary electrons left from the decay of the charged pions (Uchiyama et al. 2010, Vink 2012), and thermal bremsstrahlung contamination. For details regarding flat radio spectra of SNRs see Oni{\' c} (2013b).

\begin{figure}[h]
  \centering
  \includegraphics[width=10cm]{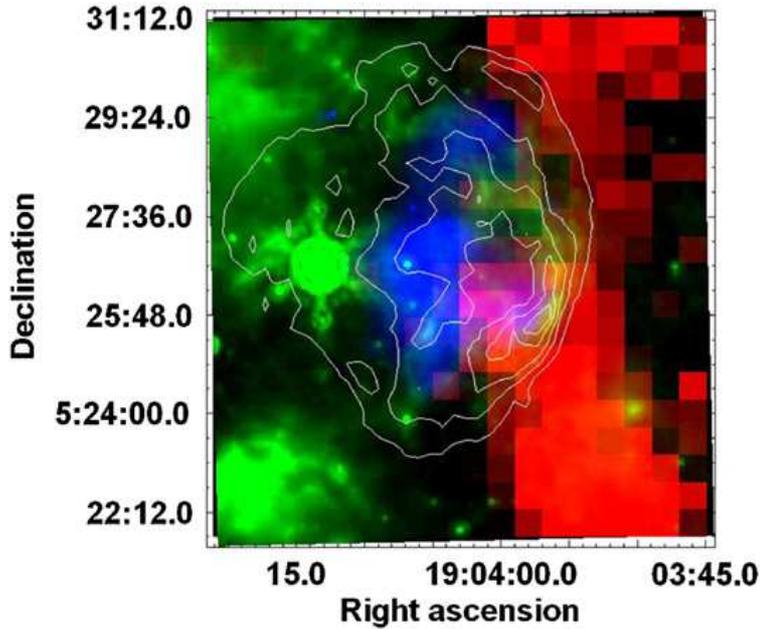}
\caption{Multiband image of composite and "radio thermally active" SNR 3C 396. The molecular gas ($^{12}$CO intensity map) is shown in red, the 24 $\mu$m
mid-IR emission in green, and 1.0 - 7.0 keV X-rays in blue. The radio contours at 1.4 GHz are also shown (Su et al. 2011).}
\end{figure}

The radio spectra of evolved SNRs can be curved. The first ideas about the significant influence of thermal bremsstrahlung in radio emission of evolved SNRs were presented in Uro{\v s}evi{\'c} (2000), and Uro{\v s}evi{\' c} et al. (2003a,b). The first attempts at identifying the concave-up forms of radio spectra of evolved SNRs can be found in Uro{\v s}evi{\' c} and Pannuti (2005 - OA184 SNR\footnote{Foster et al. (2006) concluded that OA184 is rather a Galactic HII region then an SNR.}), Uro{\v s}evi{\' c} et al. (2007) and Oni{\' c} and Uro{\v s}evi{\' c} (2008) for SNR HB3. Oni{\' c} et al. (2012) presented the possible significant contribution of the thermal bremsstrahlung emission inferred from the concave-up spectra of three Galactic SNRs: IC443, 3C391 and 3C396 (Figs. 5 and 6). All these SNRs are evolved and embedded in a high density molecular cloud environment (for details see Oni{\' c} et al. 2012, Oni{\' c} 2013a).

\begin{figure}[h]
  \centering
  \includegraphics{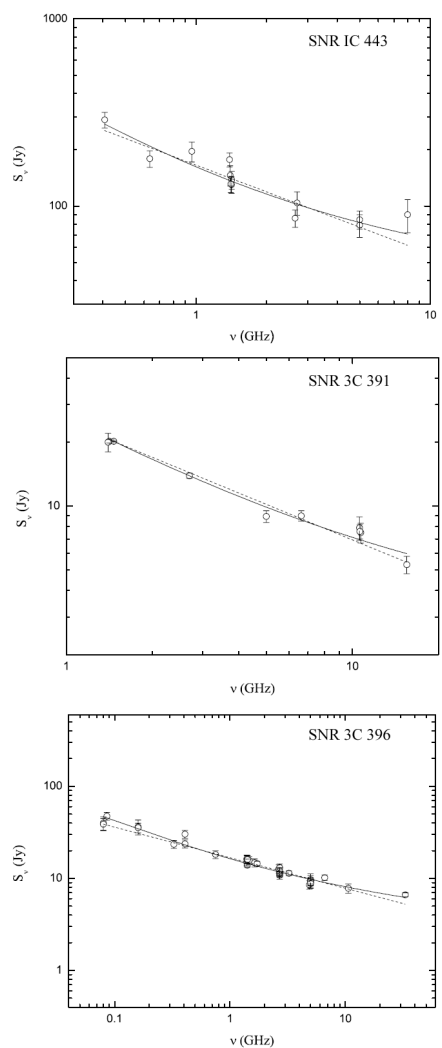}
\caption{Radio spectra of possible thermally active SNRs: IC443, 3C391 and 3C396 (Oni{\' c} et al. 2012).}
\end{figure}

By inspection and comparison of Fig. 4 and Fig. 6 it is clear that spectral curving for some young and evolved SNRs exists.  At first glance, the concave-up spectral forms for evolved SNRs are visible in Fig. 6 than for young ones (Fig. 4). On the other hand, spectra of young SNRs have more data points, thus  providing better statistics. Generally, it can be concluded that these spectra appear in similar form.

\medskip

Also, the slopes of the integrated radio continuum spectra of SNRs at higher frequencies (between 10-100 GHz) could be flatter\footnote{It results in the concave-up form of overall radio spectrum.} because of the influence of the spinning dust emission (Scaife at al. 2007). The spinning dust is responsible for a creation of a "hump" in the high-frequency radio spectrum (Draine and Lazarian 1998). For details see Oni{\' c} (2013b).

This model can not provide an extension of the concave-up form to the X-ray part of an SNR spectrum. The spectral extension into X-ray domain can be expected for all SNRs, as such as ones with significant thermal bremsstrahlung emission. Due to this, whether the thermal bremsstrahlung model or spinning dust model is responsible for a concave-up radio spectrum can be determined through X-ray observations.  The detection of overall (from radio to X-rays) concave-up spectral forms for evolved SNRs should be an interesting task for observers in near future.

\medskip

A few examples of concave-down radio spectra of evolved  SNRs can be found in recent literature. The theoretical background is given in this review, Subsection 4.2. Among the Galactic and some Large Magellanic Cloud (LMC) SNRs there is existence of spectral break and steepening at higher radio frequencies. Xiao et al (2008) identified a spectral break at 1.5 GHz for Galactic SNR S147. The low-frequency spectral index is around 0.3, and the high-frequency spectral index is around 1.2. The apparent bending was explained by effects induced by synchrotron losses, during the early phase of the SNR evolution. This would cause a bending at a rather high frequency. The subsequent expansion of S147 would shift the spectral break frequency toward about 1.5 GHz. Crawford et al. (2008) detected concave-down spectrum with steepening at frequencies around 3 GHz for evolved LMC SNR J0455-6838 (Fig. 7). The similar spectral form was detected for LMC SNR J0527-6549 (DEM L204) (Bozzetto et al. 2010), with the spectral break was detected also at around 3 GHz. For the first time Bozzetto et al. (2010) gave an explanation based on DSA theory, by the effect of coupling between synchrotron losses and observational constraints when the distant (extragalactic) emission region stayed unresolved (see Subsection 4.2 in this review). De Horta et al. (2012), Bozzetto et al. (2012), and Bozzetto et al. (2013) presented concave-down spectra of an additional three evolved LMC SNRs. They gave an explanation of bending based on the observational effects -- the shortest baseline observations as a result give missing flux because of the lack of short spacings. Additional observations should resolve whether the steepening at high frequencies originates from observational effects or effects based on theoretical predictions: probably the final explanation will be based on a combination of both. Recently, Pivato et al. (2013) detected concave-down spectrum of evolved Galactic mixed-morphology SNR HB 21, with steepening at around 6 GHz (Fig. 8). They analyzed the WMAP 7-year data in frequency range between 23 and 93 GHz, achieving the first detection of HB 21 at four frequencies in this interval. They explained the concave-down form of spectrum by the synchrotron losses in a homogeneous source of continuously injected electrons, as it is described in this review, Subsection 4.2.

\begin{figure}[h]
  \centering
  \includegraphics{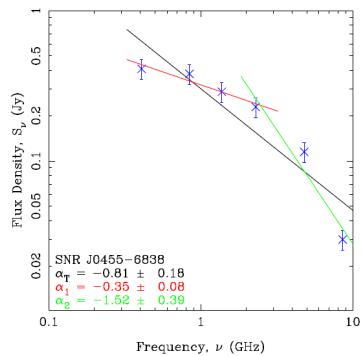}
\caption{The concave-down radio spectrum of LMC SNR J0455-6838 (Crawford et al. 2008).}
\end{figure}

\begin{figure}[h]
  \centering
  \includegraphics{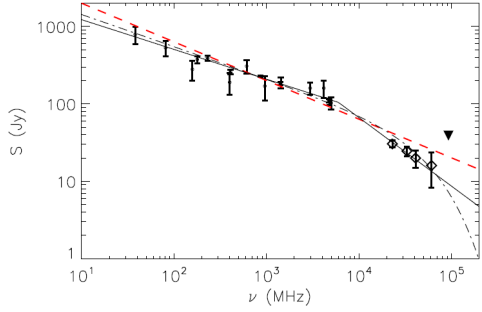}
\caption{The concave-down radio spectrum of HB21 (Pivato et al. 2013).}
\end{figure}

The low-frequency turnovers were detected in the earliest measurements made of the spectra of SNRs (e.g. Roger et al. 1973). This effect should be a result of the thermal absorption in optically thick environment or the synchrotron self-absorption, again in the opaque medium, at low radio frequencies. The corresponding theoretically derived spectral indices are $\alpha=-2$ and $\alpha=-2.5$, respectively (see Longair 2000). There are several SNRs with radio spectra which show breaks at low radio frequencies (Fig. 9). These spectra often lack data points at low radio frequencies. Due to this, the observational confirmation of which of these two processes is dominantly active could be a very demanding task. Additionally, the low-frequency absorption is mainly related to the ISM along the line-of-sight between an SNR and us as observers, for example, in the cases of the SNRs W49B (Lacey et al. 2001), CTB 80 (Castelletti and Dubner 2005), G354.4+0.0 (Roy and Pal 2013). There are a few SNRs with detected intrinsic low-frequency absorption related to the sources themselves (e.g. IC 443, 3C 391, Oni{\' c} et al. (2012, and references therein)).

\begin{figure}[h]
  \centering
  \includegraphics{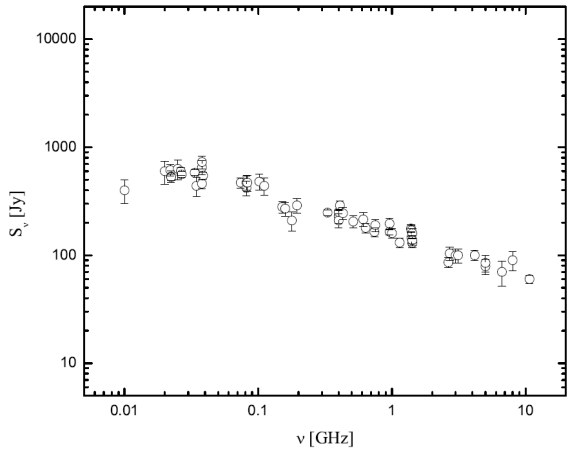}
\caption{The radio spectrum with the low-frequency turnover of SNR IC443 (Oni{\' c} 2013a).}
\end{figure}

\section{Summary}

In this review, I presented:

\begin{description}
  \item[i)]  a brief overview of fundamentals of DSA and synchrotron mechanisms.
  \item[ii)] theoretical explanations for the continuum SNR radio spectra formation. The predictions of linear and curved forms of radio spectra are presented for both young and evolved SNRs. These predictions are mainly based on DSA theory.
  \item[iii)] examples of SNR radio spectra obtained from observations which are compared with the different theoretically predicted spectral forms.
\end{description}

The theoretical predictions and some observational confirmations of different spectral forms in radio of shell-like, composite and mixed-morphology SNRs (mentioned in this review) are summarized in Table 1.

\begin{table}[h]
\caption{The theoretically predicted spectral forms and some examples for observationally obtained radio spectra of shell-like, composite and mixed-morphology SNRs.}
\begin{tabular}{l|l|l|l|l|l}
\hline\noalign{\smallskip}
\multicolumn{6}{c}{theoretical predictions} \\
\hline
\hline
&\multicolumn{2}{l}{linear radio spectra}&&\multicolumn{2}{l}{curved radio spectra}  \\ \hline
&$\alpha=0.5$&steep ($\alpha>0.5$)&flat ($\alpha<0.5$)&concave-up&concave-down \\
\hline\hline
young SNRs  & test particle & ampl. mag. field + &DSA + &non-linear&obs. effects \\
&DSA& quasi-perp. shocks&Fermi 2&DSA&\\ \hline
evolved SNRs & DSA  & test particle &DSA +&synch. + brem.&obs. effects + \\
& & DSA &Fermi 2&or spin. dust&DSA effects\\
\hline\hline\noalign{\smallskip}
\multicolumn{6}{c}{from observations} \\
\hline
\hline
&\multicolumn{2}{l}{linear radio spectra}&&\multicolumn{2}{l}{curved radio spectra}  \\ \hline
&$\alpha=0.5$&steep ($\alpha>0.5$)&flat ($\alpha<0.5$)&concave-up&concave-down \\
\hline\hline
young SNRs  & / & e.g. Cas A,&/&e.g. Tycho, &/ \\
&&G1.9+0.3&&Kepler, SN1006&\\ \hline
evolved SNRs &e.g. Monoceros and &e.g. HB3, HB9 &e.g. W28,&e.g. IC443, &e.g. S147, HB21,  \\
&Lupus loops & &Kes67, 3C434.1&3C391, 3C396&J0455-6838\\
\hline\noalign{\smallskip}\hline

\end{tabular}
\end{table}

Finally, the high-frequency and low-frequency domains of the SNR radio spectra are covered with low number of data points or no data at all. The effects of curving in spectra are connected with these ending regions of the radio domain. Due to this, the new confirmations of the theoretically predicted radio spectral features at high frequencies could be expected from archival or future observations by the upgraded ATCA (Australia Telescope Compact Array), WMAP (Wilkinson Microwave Anisotropy Probe), and ALMA (Atacama Large Millimeter/submillimeter Array)  telescopes.  Additionally, LOFAR (Low-Frequency Array) interferometer should be used for the detection of the low-frequency turnovers. In the future, SKA (Square Kilometre Array) telescope will probably give final answers for the open questions related to bending in the SNR radio spectra.



%




%

\begin{acknowledgements}
I would like to thank the referee for excellent comments and suggestions. Also,I would like to thank N. Duric, T. Pannuti, D. Oni{\' c}, B. Arbutina, and M. Pavlovi{\' c} for help in many aspects: exciting discussions, careful reviewing and editing of typescript, providing some data and references, etc. Their comments provided that the final version of this review has appeared in significantly better form.  Additionally I acknowledge the financial support of the Ministry of Education, Science, and Technological Development of the Republic of Serbia through the project No. 176005 "Emission Nebulae: Structure and Evolution".
\end{acknowledgements}




\begin{thebibliography}{}
%
%

\bibitem [\protect\citeauthoryear{authoryear}{}]{} Allen GE, Houck JC, Sturner SJ (2008) Evidence of a curved synchrotron spectrum in the supernova remnant SN 1006. Astrophys J 683:773-785

\bibitem [\protect\citeauthoryear{authoryear}{}]{} Arbutina B, Uro{\v s}evi{\' c} D, Andjeli{\' c} MM, Pavlovi{\' c} MZ, Vukoti{\' c} B (2012) modified equipartition calculation for supernova remnants. Astrophys J 746:79

\bibitem [\protect\citeauthoryear{authoryear}{}]{} Arbutina B, Uro{\v s}evi{\' c} D, Vu{\v c}eti{\' c} MM, Pavlovi{\' c} MZ, Vukoti{\' c} B (2013) Modified equipartition calculation for supernova remnants. Cases $\alpha$ = 0.5 and $\alpha$ = 1. Astrophys J 777:31

\bibitem [\protect\citeauthoryear{authoryear}{}]{} Bell AR (1978a) The acceleration of cosmic rays in shock fronts. I. Mon Not R Astron Soc 182:147-156

\bibitem [\protect\citeauthoryear{authoryear}{}]{} Bell AR (1978b) The acceleration of cosmic rays in shock fronts. II. Mon Not R Astron Soc 182:443-455

\bibitem [\protect\citeauthoryear{authoryear}{}]{} Bell AR (2004) Turbulent amplification of magnetic field and diffusive shock acceleration of cosmic rays. Mon Not R Astron Soc 353:550-558

\bibitem [\protect\citeauthoryear{authoryear}{}]{} Bell AR (2005) The interaction of cosmic rays and magnetized plasma. Mon Not R Astron Soc 358:181-187

\bibitem [\protect\citeauthoryear{authoryear}{}]{} Bell AR, Schure KM, Reville B (2011) Cosmic ray acceleration at oblique shocks. Mon Not R Astron Soc 418:1208-1216

\bibitem [\protect\citeauthoryear{authoryear}{}]{} Berezhko EG, V{\"o}lk HJ (2004) The theory of synchrotron emission from supernova remnants. Astron Astrophys 427:525–536

\bibitem [\protect\citeauthoryear{authoryear}{}]{} Blandford RD, Ostriker JP (1978) Particle acceleration by astrophysical shocks. Astrophys J 221:L29-L32

\bibitem [\protect\citeauthoryear{authoryear}{}]{} Bozzetto LM, Filipovi{\' c} MD, Crawford EJ, Boji{\v c}i{\' c} IS, Payne JL,
Mendik A, Wardlaw B, De Horta AY (2010) Multifrequency radio observations of a SNR in the LMC. The case of SNR J0527-6549 (DEM L204). Serb Astron J 181:43-49

\bibitem [\protect\citeauthoryear{authoryear}{}]{} Bozzetto LM, Filipovi{\' c} MD, Crawford EJ, Payne JL, De Horta AY, Stupar M (2012) Radio continuum observations of LMC SNR J0550-6823. Rev Mex Astron Astrophys 48:41–46

\bibitem [\protect\citeauthoryear{authoryear}{}]{} Bozzetto LM, Filipovi{\' c} MD, Crawford EJ, Sasaki M, Maggi P, Haberl F, Uro{\v s}evi{\' c} D, Payne JL, De Horta AY, Stupar M, Gruendl R, Dickel J (2013) Multifrequency study of SNR J0533-7202, a new supernova remnant in the LMC.  Mon Not R Astron Soc 432:2177–2181

\bibitem [\protect\citeauthoryear{authoryear}{}]{} Braude SYa, Megn AV, Ryabov BP, Zhouck IN (1970) The spectra of some discrete radio sources in 10-5000 MHz frequency range. Astrophys Space Sci 8:275-322

\bibitem [\protect\citeauthoryear{authoryear}{}]{} Castelletti G, Dubner G (2005) A multi-frequency study of the spectral index distribution in the SNR CTB 80.  Astron Astrophys 440:171-177

\bibitem [\protect\citeauthoryear{authoryear}{}]{} Crawford EJ, Filipovi{\' c} MD, De Horta AY, Stootman FH, Payne JL (2008) Radio-continuum study of the supernova remnants in the Large Magellanic Cloud - an SNR with a highly polarised breakout region - SNR J0455-6838. Serb Astron J 177:61-66

\bibitem [\protect\citeauthoryear{authoryear}{}]{} Cummings AC (1973) A study of cosmic ray positron and electron spectra in interplanetary
and interstellar space and the solar modulation of cosmic rays.  Ph.D. thesis, Caltech

\bibitem [\protect\citeauthoryear{authoryear}{}]{} De Horta AY, Filipovi{\' c} MD, Bozzetto LM, Maggi P, Haberl F, Crawford EJ, Sasaki M,  Uro{\v s}evi{\' c} D, Pietsch W, Gruendl R, Dickel J, Tothill NFH, Chu Y-H, Payne JL, Collier JD  (2012) Multi-frequency study of supernova remnants in the Large Magellanic Cloud. The case of LMC SNR J0530–7007. Astron Astrophys 540:25

\bibitem [\protect\citeauthoryear{authoryear}{}]{} Draine BT, Lazarian A (1998) Diffuse galactic emission from spinning dust grains. Astrophys J 494:L19-L22

\bibitem [\protect\citeauthoryear{authoryear}{}]{} Drury LO'C (1983) An introduction to the theory of diffusive shock acceleration of energetic particles in tenuous plasmas. Rep Prog Phys 46:973-1027

\bibitem [\protect\citeauthoryear{authoryear}{}]{} Drury LO'C, V{\" o}lk HJ (1981) Hydromagnetic shock structure in the presence of cosmic rays. Astron Astrophys 248:344-351

 \bibitem [\protect\citeauthoryear{authoryear}{}]{} Eichler D (1984) On the theory of cosmic-ray-mediated shocks with variable compression ratio. Astrophys J 277:429-434

\bibitem [\protect\citeauthoryear{authoryear}{}]{} Fermi E (1949) On the origin of the cosmic radiation. Phys Rev 75:1169-1174

\bibitem [\protect\citeauthoryear{authoryear}{}]{} Foster T, Kothes R, Sun XH, Reich W, Han JL (2006) $10^{51}$ erg less: the Galactic H II region OA 184. Astron Astrophys 454:517-526

\bibitem [\protect\citeauthoryear{authoryear}{}]{} Gao XY, Han JL, Reich W, Reich P, Sun XH, Xiao L(2011) A Sino-German $\lambda$6 cm polarization survey of the Galactic plane. V. Large supernova remnants. Astron Astrophys 529:159

\bibitem [\protect\citeauthoryear{authoryear}{}]{} Ginzburg VL, Syrovatskii SI (1965) Cosmic magnetobremsstrahlung (synchrotron radiation). Annu Rev Astron Astrophys  3:297-350

\bibitem [\protect\citeauthoryear{authoryear}{}]{} Green DA (2009) A revised Galactic supernova remnant catalogue. Bull Astron Soc India 37:45-61

\bibitem [\protect\citeauthoryear{authoryear}{}]{} Guseinov OH, Ankay A, Tagieva SO (2003) Observational data on galactic supernova remnants: I. The supernova remnants within l=0-90 degrees. Serb Astron J 167:93-110

\bibitem [\protect\citeauthoryear{authoryear}{}]{} Guseinov OH, Ankay A, Tagieva SO (2004a) Observational data on galactic supernova remnants: II. The supernova remnants within l=90-270 degrees. Serb Astron J 168:55-69

\bibitem [\protect\citeauthoryear{authoryear}{}]{} Guseinov OH, Ankay A, Tagieva SO (2004b) Observational data on galactic supernova remnants: III. The supernova remnants within l=270-360 degrees. Serb Astron J 169:65-82

\bibitem [\protect\citeauthoryear{authoryear}{}]{} Heavens AF, Meisenheimer K (1987) Particle acceleration in extragalactic sources - the role of synchrotron losses in determining the spectrum. Mon Not R Astron Soc 225:335-353

\bibitem [\protect\citeauthoryear{authoryear}{}]{} Jiang ZJ, Zhang L, Fang J (2013) Spectral evolution of accelerated particles in supernova remnants. Mon Not R Astron Soc 433:1271-1275

\bibitem [\protect\citeauthoryear{authoryear}{}]{} Jones TJ, Rudnick L, DeLaney T, Bowden J (2003) The identification of infrared synchrotron radiation from Cassiopeia A. Astrophys J 587:227-234

 \bibitem [\protect\citeauthoryear{authoryear}{}]{}  Jones TW, Rudnick L., Jun B-Il, Borkowski KJ,
Dubner G., Frail DA, Kang H, Kassim NE, McCray R (1998) $10^{51}$ ergs: the evolution of shell supernova remnants. Publ Astron Soc Pacific 110:125-151

\bibitem [\protect\citeauthoryear{authoryear}{}]{} Lacey CK, Lazio T, Joseph W, Kassim NE, Duric N, Briggs DS, Dyer KK (2001) Spatially resolved thermal continuum absorption against supernova remnant W49B. Astrophys J 559:954-962

 \bibitem [\protect\citeauthoryear{authoryear}{}]{} Lequeux J (2005) The interstellar medium, with the collaboration of E Falgarone and C Ryter. Springer-Verlag, Berlin, Heidelberg

\bibitem [\protect\citeauthoryear{authoryear}{}]{} Longair MS (2000) High energy astrophysics. Volume 2. Stars, the Galaxy and the interstellar medium. Cambridge University Press, Cambridge (UK)

\bibitem [\protect\citeauthoryear{authoryear}{}]{} McKee CF, Ostriker JP (1977) A theory of the interstellar medium - three components regulated by supernova explosions in an inhomogeneous substrate. Astrophys J 218:148-169

\bibitem [\protect\citeauthoryear{authoryear}{}]{} Oni{\' c} D (2013a) Thermal radiation of supernova remnants in radio domain. Ph.D. thesis, University of Belgrade

\bibitem [\protect\citeauthoryear{authoryear}{}]{} Oni{\' c} D (2013b) On the supernova remnants with flat radio spectra.  Astrophys Space Sci 346:3-13

\bibitem [\protect\citeauthoryear{authoryear}{}]{} Oni{\' c} D, Uro{\v s}evi{\' c} D (2008) The analysis of the possible thermal emission at radio frequencies from an evolved supernova remnant HB 3 (G132.7+1.3): revisited.  Serb astron J 177:67-71

\bibitem [\protect\citeauthoryear{authoryear}{}]{} Oni{\' c} D, Uro{\v s}evi{\' c} D, Arbutina B, Leahy D (2012) On the existence of "radio thermally active" Galactic supernova remnants.  Astrophys J 756:61

\bibitem [\protect\citeauthoryear{authoryear}{}]{} Ostrowski M (1999) Supernova remnants in molecular clouds: on cosmic ray electron spectra. Astron Astrophys 345:256–258

\bibitem [\protect\citeauthoryear{authoryear}{}]{} Pavlovi{\' c} MZ, Uro{\v s}evi{\' c} D, Vukoti{\' c} B, Arbutina B, G{\" o}ker {\" U}D (2013) The radio surface-brightness-to-diameter relation for Galactic supernova remnants: sample selection and robust analysis with various fitting offsets.  Astrophys J Suppl Ser 204:4

\bibitem [\protect\citeauthoryear{authoryear}{}]{} Pivato G, Hewitt J, Tibaldo L, Acero F, Ballet J, Brandt TJ, de Palma F, Giordano F, Janssen GH, Johannesson G, Smith DA (2013) Fermi LAT and WMAP observations of the supernova remnant HB 21. Astrophys J 779:179

\bibitem [\protect\citeauthoryear{authoryear}{}]{} Reynolds SP (2008) Supernova remnants at high energy. Annu Rev Astron Astrophys 46:89-126

\bibitem [\protect\citeauthoryear{authoryear}{}]{} Reynolds SP, Ellison DC (1992) Electron acceleration in Tycho's and Kepler's supernova remnants - spectral evidence of Fermi shock acceleration. Astrophys J 399:L75-L78

\bibitem [\protect\citeauthoryear{authoryear}{}]{} Reynolds SP, Gaensler BM, Bocchino F (2012) Magnetic fields in supernova remnants and pulsar-wind nebulae. Space Sci Rev 166:231-261

\bibitem [\protect\citeauthoryear{authoryear}{}]{} Rho J, Petre R (1998) Mixed-morphology supernova remnants. Astrophys J 503:L167-L170

\bibitem [\protect\citeauthoryear{authoryear}{}]{} Roger RS, Bridle AH, Costain CH (1973) The low-frequency spectra of nonthermal radio sources. Astron J 78:1030-1057

\bibitem [\protect\citeauthoryear{authoryear}{}]{} Roy S, Pal S (2013) Discovery of the Small-diameter, Young Supernova Remnant G354.4+0.0. Astrophys J 774:150

\bibitem [\protect\citeauthoryear{authoryear}{}]{} Scaife A, Green DA, Battye RA, Davies RD, Davis RJ, Dickinson C, Franzen T, Génova-Santos R, Grainge K, Hafez YA, Hobson MP, Lasenby A, Pooley GG, Rajguru N, Rebolo R, Rubino-Martin JA, Saunders RDE, Scott PF, Titterington D, Waldram E, Watson RA (2007) Constraints on spinning dust towards Galactic targets with the Very Small Array: a tentative detection of excess microwave emission towards 3C396.  Mon Not R Astron Soc 377:L69–L73

\bibitem [\protect\citeauthoryear{authoryear}{}]{} Schlickeiser R, F{\" u}rst E (1989) The origin of flat radio spectra in shell-type supernova remnants. Astron Astrophys 219:192-194

\bibitem [\protect\citeauthoryear{authoryear}{}]{} Su Y, Che Y, Yang J, Koo B-C, Zhou X, Lu D-R, Jeong I-G, DeLaney T (2011) Molecular environment and thermal X-ray spectroscopy of the semicircular young composite supernova remnant 3C 396. Astrophys J 727:43


\bibitem [\protect\citeauthoryear{authoryear}{}]{} Sun XH, Reich P, Reich W, Xiao L, Gao XY, Han JL (2011) A Sino-German $\lambda$6 cm polarization survey of the Galactic plane. VII. Small supernova remnants. Astron Astrophys 536:83


\bibitem [\protect\citeauthoryear{authoryear}{}]{} Uchiyama Y, Blandford RD, Funk S, Tajima H, Tanaka T (2010) Gamma-ray emission from crushed clouds in supernova remnants. Astrophys J 723:L122-L126

\bibitem [\protect\citeauthoryear{authoryear}{}]{} Uro{\v s}evi{\'c} D (2000) The $\Sigma-D$ relation as an indicator of radio loop origin. Ph.D. thesis, University of Belgrade

\bibitem [\protect\citeauthoryear{authoryear}{}]{} Uro{\v s}evi{\' c} D, Duric N, Pannuti TG (2003a) A modified theoretical $\Sigma-D$ relation for supernova remnants: I. The case of constant temperature within the supernova remnant. Serb Astron J 166:61-66

\bibitem [\protect\citeauthoryear{authoryear}{}]{} Uro{\v s}evi{\' c} D, Duric N, Pannuti TG (2003b) A modified theoretical $\Sigma-D$ relation for supernova remnants: II. The case of variable temperature within the supernova remnant. Serb Astron J 166:67-70

\bibitem [\protect\citeauthoryear{authoryear}{}]{} Uro{\v s}evi{\' c} D, Pannuti TG (2005) Thermal emission at radio frequencies from supernova remnants and a modified theoretical $\Sigma-D$ relation. Astropart Phys 23:577-587

\bibitem [\protect\citeauthoryear{authoryear}{}]{} Uro{\v s}evi{\' c} D, Pannuti TG, Leahy D (2007) An analysis of the broadband (22-3900 MHz) radio spectrum of HB 3 (G132.7+1.3): the detection of thermal radio emission from an evolved supernova remnant? Astrophys J 655:L41-L44

\bibitem [\protect\citeauthoryear{authoryear}{}]{} Vink J (2012) Supernova remnants: the X-ray perspective. Astron Astrophys Rev 20:49

\bibitem [\protect\citeauthoryear{authoryear}{}]{} Vink J, Bleeker J, van der Heyden K, Bykov A, Bamba A, Yamazak R (2006) The X-ray synchrotron emission of RCW 86 and the implications for its age. Astrophys J 648:L33-L37

\bibitem [\protect\citeauthoryear{authoryear}{}]{} Woltjer L (1972) Supernova remnants. Annu Rev Astron Astrophys 10:129-158

\bibitem [\protect\citeauthoryear{authoryear}{}]{} Xiao L, F{\" u}rst E, Reich W, Han JL (2008) Radio spectral properties and the magnetic field of the SNR S147. Astron Astrophys 482:783-792






\end{thebibliography}


\end{document}